\documentclass{article}
\usepackage{arxiv}
\usepackage{graphicx}
\usepackage{subcaption}
\usepackage{amsmath}
\usepackage{multirow}
\usepackage{amsfonts}
\usepackage{amsmath}
\usepackage[T1]{fontenc}    % use 8-bit T1 fonts
\usepackage{hyperref}       % hyperlinks
\usepackage{url}            % simple URL typesetting
\usepackage{booktabs}       % professional-quality tables
\usepackage{amsfonts}       % blackboard math symbols
\usepackage{nicefrac}       % compact symbols for 1/2, etc.
\usepackage{microtype}      % microtypography
\usepackage{lipsum}
\usepackage{tikz}
\usepackage{verbatim}
\usepackage{textcomp}
\usepackage[ansinew]{inputenc}

\title{Damage Detection in Bridge Structures: An Edge Computing Approach}

\author{
  Rahul Kumar Verma \\
  Wireless Sensor Networks Laboratory\\
  ABV-Indian Institute of Information Technology and Management, Gwalior\\
  India\\
  \texttt{rahulkv@ieee.org} \\
   \And
 K. K. Pattanaik \\
  Wireless Sensor Networks Laboratory\\
  ABV-Indian Institute of Information Technology and Management, Gwalior\\
  India\\
  \texttt{kkpatnaik@iiitm.ac.in} \\
  \And
 P. B. R. Dissanayake \\
  Department of Civil Engineering\\
  University of Peradeniya\\
  Sri Lanka\\
  \And
  A. J. Dammika \\
  Department of Civil Engineering\\
  University of Peradeniya\\
  Sri Lanka\\
  \And
  H. A. D. Samith Buddika \\
  Department of Civil Engineering\\
  University of Peradeniya\\
  Sri Lanka\\
  \And
  Mosbeh R. Kaloop \\
  Public Works and Civil Engineering Department\\
  Mansoura University\\
  Egypt\\
}  
  
\begin{document}
\maketitle

\begin{abstract}
Wireless sensor network (WSN) based SHM systems have shown significant improvement as compared to traditional wired-SHM systems in terms of cost, accuracy, and reliability of the monitoring. However, due to the resource-constrained nature of the sensor nodes, it is a challenge to process a large amount of sensed vibration data in real-time. Existing mechanisms of data processing are centralized and use cloud or remote servers to analyze the data to characterize the state of the bridge, i.e., healthy or damaged. These methods are feasible for wired-SHM systems, however, transmitting huge data-sets in WSNs has been found to be arduous. In this paper, we propose a mechanism named as ``in-network damage detection on edge (INDDE)" which extracts the statistical features from raw acceleration measurements corresponding to the healthy condition of the bridge and use them to train a probabilistic model, i.e., estimating the probability density function (PDF) of multivariate Gaussian distribution. The trained model helps to identify the anomalous behaviour of the new data points collected from the unknown condition of the bridge in real-time. Each edge device classifies the condition of the bridge as either ``healthy" or ``damaged" around its deployment region depending on their respective trained model. Experimentation results showcase a promising $\approx 96-100\%$ damage detection accuracy with the advantage of no data transmission from sensor nodes to the cloud for processing.
\end{abstract}

% keywords can be removed

\section{Introduction}
With the advent of Internet of Things (IoT) technologies, civil structures, such as bridges, buildings, wind turbines, dams, etc., are embedded with the smart sensors, forming a wireless sensor network (WSN), for continuous monitoring of their structural health. The idea behind structural health monitoring (SHM) is to collect the data concerning to  various physical and chemical parameters (such as temperature, humidity, pH of the concrete, corrosion, etc.), and mechanical parameters (such as strains, deflections, vibrations, etc.) from sensor nodes strategically placed over the structure to assess structural condition \cite{shm_survey}. In the past few decades, deterioration of bridges has received attention of researchers in the field of structural engineering and computer science to come up with the solutions for long-term monitoring for bridges to detect damages.

Bridges are essential for any surface transportation network and play a significant role in connecting various separated locations. Bridges are subjected to varying loads (wind, vehicles, pedestrian, etc.) and environmental conditions, which makes them prone to damage and faster health deterioration. Thus, the safety and real-time monitoring of the bridges are of utmost importance for a trouble-free transportation network. Furthermore, real-time monitoring of the bridges enables to take necessary actions on the identification of structural deformation under a load of high traffic, or after a catastrophic event, such as the earthquake, flood, terrorist's attack, etc. In the case of bridge damage detection, acceleration data are used mostly because acceleration is directly influenced by the external forces and reflects the structural damages directly \cite{kaloop4}\cite{vibration}. Traditionally, modal parameter based techniques are used which depend on characteristics of the structure, such as natural frequencies, mode shapes, damping ratios, etc., identified from acceleration data and compared between the healthy and current states in order to characterise the damage(s) in the structure. 

The existing damage detection mechanisms are predominantly centralized as they require transmission of data from all the deployed sensors to a single processing unit in order to assess the structural integrity. These approaches are relatively economical for wired-SHM systems; whereas for WSN-based SHM systems, the transmission of enormous amount of data is very challenging due to constrained resources of sensor nodes \cite{intro}. The use of centralized architecture for SHM is not suitable in context of resource-constrained WSNs since sensors collect vibration measurements using accelerometers at high frequency during a time period, thus, contain a sequence of over thousands of data points to be transmitted \cite{shm-cao}\cite{shm_framework}\cite{shm_fog}. Wireless transmission costs more energy than the local processing, thus poses several challenges for battery-powered wireless nodes \cite{gda}. Moreover, centralized systems are inadequate for real-time detection of the damages due to prolonged time needed for collecting the data and analyzing it off-line. 

A feasible approach to overcome this architectural restriction is to perform damage detection process inside the network, which further reduces the data transmissions. Various in-network data processing mechanisms \cite{gda}\cite{god}\cite{edgemining} have been proved energy-efficient in different application domains of WSNs. The adoption of WSNs for SHM brings several advantages when compared to other approaches, such as in-situ inspections and wired-SHM systems. One advantage is the hassle-free deployment, and another is local processing capabilities of sensor nodes to perform computational tasks as well as making intelligent decisions upon collected data. Performing online processing within the network (i.e., in-network processing) provides responses more quickly concerning to an event (damage) than the solutions based on centralized or remote control centres. It is noteworthy to mention that by taking the damage detection process inside the network, new challenges arise.

A key challenge in WSN-based SHM systems is to minimize the data transmission to conserve energy of sensor nodes while meeting the application's required detection accuracy. Some of the existing modal-based SHM algorithms, e.g., eigensystem realization algorithm (ERA), are made distributed but they suffer from high network traffic, energy consumption, and latency during the decision-making process due to frequent data exchanges among sensor nodes \cite{shm-cao}. Furthermore, traditional damage detection mechanisms use dynamic characteristics of the structure, such as natural frequency, modal shapes, modal damping, etc., which are not very convenient for real-time online damage detection and difficult to be extracted especially when the structure is excited by low-frequency inputs \cite{shm-stat}\cite{decentralized}. Therefore, a decentralized and computationally lightweight real-time damage detection mechanism is required that would reduce the number of transmissions in the network as well as the workload concentrated at the central server. The interpretation of raw vibration data using statistical methods and computational intelligence techniques have been considered as a viable solution for damage detection in WSN-based SHM \cite{kaloop}\cite{ML}\cite{distributed}\cite{santosML}. The main advantage of using raw vibration responses is that the original data can be processed directly on the sensor nodes, which overcome the effects caused by the modal parameters based damage detection techniques.

Taking a cue from above discussion, this work proposes a computationally lightweight real-time damage detection mechanism, named as ``In-Network Damage Detection on Edge (INDDE)". INDDE is a decentralized mechanism which enables each edge device to perform damage detection process locally instead of sending acceleration data to a central entity (e.g., base station or cloud). INDDE extracts statistical features from acceleration data to train the model at each sensor node corresponding to the healthy condition of the bridge. Acceleration measurements of the bridge during its normal operational state are inexpensive and easy to obtain. Conversely, measurements of damaged bridge would require the destruction of bridge in all possible ways. Therefore, it is difficult to obtain the data corresponding to damaged condition of the bridge, which limits the applicability of various machine learning approaches in such applicative scenarios. In this paper, we formulate the damage detection problem as \textit{``anomaly detection"} in which the probability density function (PDF) of multivariate Gaussian distribution is estimated using statistical features representing healthy condition of the bridge. INDDE fits the statistical damage sensitive features into the probability distribution model, which eliminates the need for data transmission and require minimal computational time and power. Each sensor node use their respective PDFs to test the abnormality, i.e., damage, in their vicinity from the data observed in unknown condition of the bridge. INDDE is computationally inexpensive statistical analysis approach, which can be deployed on any single-board computer (such as Arduino, Raspberry Pi, etc.) and sensor motes (such as TelosB, XBee, etc.). It helps in detecting the damage in real-time through on-node analysis of time-domain acceleration data.
The contributions of this paper are outlined below.

\begin{enumerate}
\item We leverage the fundamentals of edge computing paradigm and propose a real-time on-node damage detection mechanism (INDDE) for WSN-based SHM. Each sensor node, executing INDDE, transmit only final result (i.e., ``healthy" or ``damaged") to the base station at a specified time-interval. Thus, INDDE significantly reduces the network traffic leading to reduced energy consumption as well as delay.

\item INDDE estimates the PDF of multivariate Gaussian distribution from statistical features of vibration data measured in healthy condition of the bridge. This estimated PDF is used to classify the unknown condition of the bridge as either healthy or damaged.

\item Experimental evaluation of INDDE performed on two  case studies shows promising 96-100\% accuracy of correctly identifying the condition of the bridge. 
\end{enumerate}

The remainder of this paper is structured as follows: Section \ref{rw} overviews related research. In Section \ref{p@e}, the overall architecture of the INDDE mechanism is explained followed by two case studies in Section \ref{casestudy}: (a) In the first case study, we use acceleration data obtained from Younghe bridge to evaluate the proposed mechanism. (b) In the second case study, we acquire vibration data from a lab-scale testbed containing a fully fixed steel beam subjected to progressive multiple damages with different severity. Finally, Section \ref{conc} concludes the paper.

\section{Related Research}
\label{rw}
One of the challenging characteristics of WSN-based SHM systems is the processing of large amount of data generated by sensor nodes for damage detection. In recent years, a lot of efforts have been made into vibration based damage detection mechanisms which utilize the vibration response of the monitored structure to assess its health; and identify and locate structural damage(s). In this section,  the taxonomy of damage detection mechanisms in WSN-based SHM is dicsussed (see Figure \ref{taxo}). Subsequently, the research efforts made in the direction of damage detection using vibration data is discussed (see TABLE \ref{related}).

\begin{table*}[h!]
\centering
\renewcommand{\arraystretch}{1.2}
\caption{Comparison of damage detection mechanisms in WSN-based SHM systems}
\label{related}
\begin{tabular}{p{3.5cm} p{6.5cm} p{2.5cm} p{3cm}}
\hline
\multicolumn{1}{l}{Reference} & Mechanism & Architecture & Technique    \\ \hline

Liu et. al. \cite{liu2012distributed} &Distributed ERA &Distributed &Modal-based\\

Forstner and Wenzel \cite{forstner} &Data mining  &Centralized &Modal-based \\

Ling et. al \cite{localized}   &Auto-regressive (AR) and Autoregressive with exogenous input (ARX)  &Localized  &Non-Modal-based \\

O'Connor et. al. \cite{o2017long} &Statistical Process Control (SPC) and Gaussian Process Regression (GPR) &Centralized &Non-Modal-based  \\

Santos et. al. \cite{santosML} &Machine learning &Centralized &Non-Modal-based \\

Abdeljaber et. al. \cite{2017real} &1D - Convolutional Neural Network (CNN)		&Localized	&Non-Modal-based \\

Anaissi et. al. \cite{onesvm}	&One Class Support Vector Machine (OCSVM)	&Localized	&Non-Modal-based \\
\hline
\end{tabular}
\end{table*}

\begin{figure*}
\centering
\includegraphics[scale=0.7]{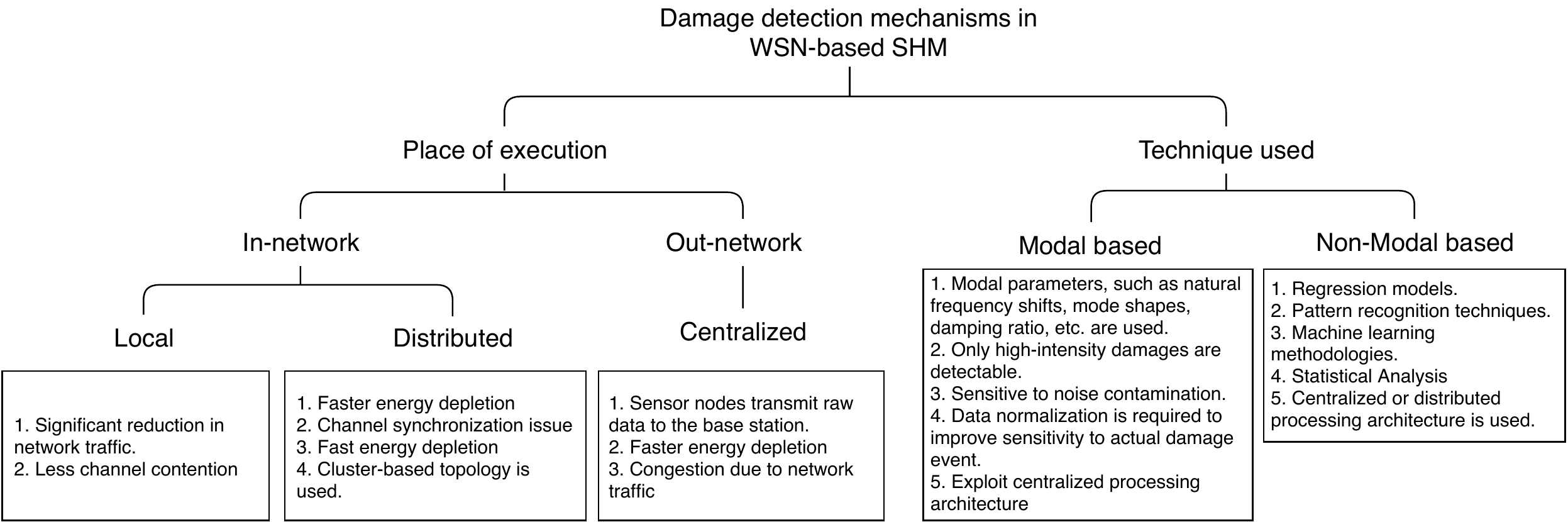}
\label{taxo}
\caption{Taxonomy of damage detection mechanisms in WSN-based SHM systems}
\label{taxo}
\end{figure*}

\subsection{On place of execution}
The damage detection mechanisms in WSN-based SHM systems can be classified into two categories based on their place of execution, i.e., out-network (centralized) and in-network (local and distributed). Majority of the existing SHM systems generally focus on off-line damage detection by collecting raw sensor data at the cloud or remote server for processing \cite{summary}. Comisu et. al. \cite{SHMsystem} proposed a centralized bridge damage detection mechanism based on 3D finite element (FE) model simulating the original structure of the bridge. The FE model is used to establish the critical thresholds for various physical and dynamics parameters that are sent wirelessly to the system for processing and detecting the damage. Chang and Kim \cite{modal} proposed centralized damage detection mechanism by using vibration data for Experimental Modal Analysis (EMA) and Operational Modal Analysis (OMA). In order to perform modal parameter estimation, battery-powered sensor nodes transmit their data to the base station, which causes high energy consumption. Moreover, the identification of modal properties for damage detection is drastically affected by environmental conditions \cite{rageh}\cite{hu-comparison}. These methods are feasible for wired-SHM systems; however, transmission of huge amount of data in WSNs has been found to be expensive for resource-constrained sensor nodes \cite{shm_survey}. Therefore, data-driven approaches for SHM are preferred and adopted to process the vibration data in decentralized (on-node or distributed) manner. In the literature, various SHM mechanisms based on statistical damage features are found better than those based on modal features  \cite{o2017long}\cite{li2010statistical}\cite{ML}, but they are centralized and causing overhead due to data transmissions and extensive computations.

Recently, hierarchical processing has been used intensively in WSNs for SHM. Alkindi et. al. \cite{intro} proposed a distributed OMA technique based on the  frequency domain decomposition (FDD) for WSN-based SHM. In this mechanism, each sensor node perform FFT locally and transmit reduced FFT result to a central node for estimating the modal parameters. Hackmann et. al. \cite{hackmann} proposed another distributed FDD mechanism to further reduce the data transmission in the network. In this mechanism, each sensor node perform FFT and construct power spectral density (PSD) for peak picking locally from PSD. The frequency domain data from selected peaks are transmitted to the central node for further FDD process. Cho et. al. \cite{dist-ssi} exploits the cluster-based architecture of WSN and propose a distributed variant of Stochastic Subspace Identification (SSI) algorithm for OMA. In this mechanism, each cluster-head execute SSI algorithm and transmit the partial modal parameters (i.e., natural frequencies and partial mode shape) to the sink. Sink combines the received outcomes from each cluster-head together to obtain global natural frequencies and mode shape. Although cluster-tree topology in WSN-based SHM outperforms centralized approach, it results in excessive computation at the cluster-level. Furthermore, cluster-based SHM is resource consuming, where cluster heads perform a lot of computation for modal analysis or damage identification. Fang et. al. \cite{cluster-shm} proposed cluster optimization technique for WSN-based SHM systems to achieve scalable and efficient sensor deployment in addition to achieving a trade-off between the energy consumption and damage detection accuracy. Liu et. al. \cite{liu2012distributed} proposed a distributed version of eigen-system realization algorithm (ERA) for WSN-based SHM. Wang et. al. \cite{wang2007} proposed a distributed processing framework for damage detection in WSN-based SHM, which consists of two algorithms, namely distributed damage index detection (DDID) running on each sensor node and collaborative damage event detection (CDED) which collaborates information from different sensor nodes to avoid false damage detection.

\subsection{On techniques used}
The physical characteristics of the bridges (e.g., stiffness, mass or damping) change with the occurrence of damage. Therefore, one of the key factors in a successful implementation of any vibration-based damage detection technique is the appropriate selection of damage sensitive features from the measured vibration response. Traditionally, modal parameters (e.g. natural frequencies, damping, and mode shapes), and their derivatives, such as modal strain energy and flexibility matrix are adopted to identify the damage in the bridge. However, various shortcomings in modal-based techniques make them less suitable for practical use-case scenarios \cite{khoa2018structural}. Such mechanisms are time consuming and expensive since they depend on the experiments to measure mode shapes and damping, thus can not be adopted for online damage detection. Moreover, it is difficult to establish a universal methodology for different structures because of distinct properties of each individual structure.

In recent years, numerous non-modal vibration-based damage detection mechanisms (e.g., time-domain analysis) have also attracted attention. In such techniques, damage is identified by comparing the current statistical features with its baseline. Various autoregressive models have also been used for bridge damage detection \cite{yao2012autoregressive}\cite{roy2015arx} in which features are either based on the residues between the prediction from an autoregressive model and the actual measured time history at each time interval, or they are simply based on autoregressive model coefficients. Machine learning based approaches \cite{ML}\cite{onesvm}\cite{kiranyaz} have also been used widely for damage detection wherein new measurements are classified through trained model to detect any structural change. These mechanisms are solely rely on raw vibration data that has to be  transformed into meaningful information to train the model. These models can overcome the problems associated with environmental and operational variability in SHM since data measured in different conditions are employed to train the models, which is not possible in modal-based techniques. Zhou et al. \cite{zhou2015} proposed a method based on wavelet package analysis, posterior probability support vector machines, and the Dempster-Shafer evidence theory to identify the structural damage. The method was tested on a 4-storey benchmark building structure and was proved to avoid the effect of any sensor failure and give more robust results. Recent studies have shown that deep learning based mechanisms can detect and locate structural damage directly from the raw vibration signals with high accuracy. Yu et. al.  \cite{yu2019novel} and Khodabandehlou et. al. \cite{2019vibration} designed and trained a convolutional neural network (CNN) with vibration data generated in different conditions ranging from ``no damage" to ``extreme damage" to locate and quantify the structural damages. Abdeljaber et al. \cite{2017real} used 1D CNN for the first time in vibration-based structural health monitoring. In an experimental study by Avci et al. \cite{avci2018wireless}, 1D CNN-based method was integrated with WSN to detect and localize the damage directly from ambient vibration response of the structure. In practice, data corresponding to damaged condition are often unavailable which restricts the use of supervised learning techniques in real case scenarios. Therefore, Anaissi et. al. \cite{onesvm} used one-class support vector machine (OCSVM) for damage detection. OCSVM uses data from healthy condition of the bridge to train the model and find whether data from unknown condition of the bridge is alike or not as the training data? If newly encountered data is too different from the trained model, it is classified as damage.

Based on the previously discussed approaches, it can be outlined that the place of data processing as well as techniques used for damage detection play a crucial role in WSN-based SHM systems. The out-network processing of data can accurately detect the damage but at the cost of high amount of traffic and increased energy consumption of WSN. On the other hand, lightweight on-node data processing is beneficial for reducing the communication and computation overhead of sensor nodes while detecting the damage. Keeping this in mind, an on-node damage detection mechanism (INDDE) is proposed in this paper, which involves no message exchanges and uses statistical approach for damage detection using time-domain acceleration data.

\section{Processing@Edge}
\label{p@e}
\begin{figure}[ht!]
\centering
\includegraphics[scale=0.7]{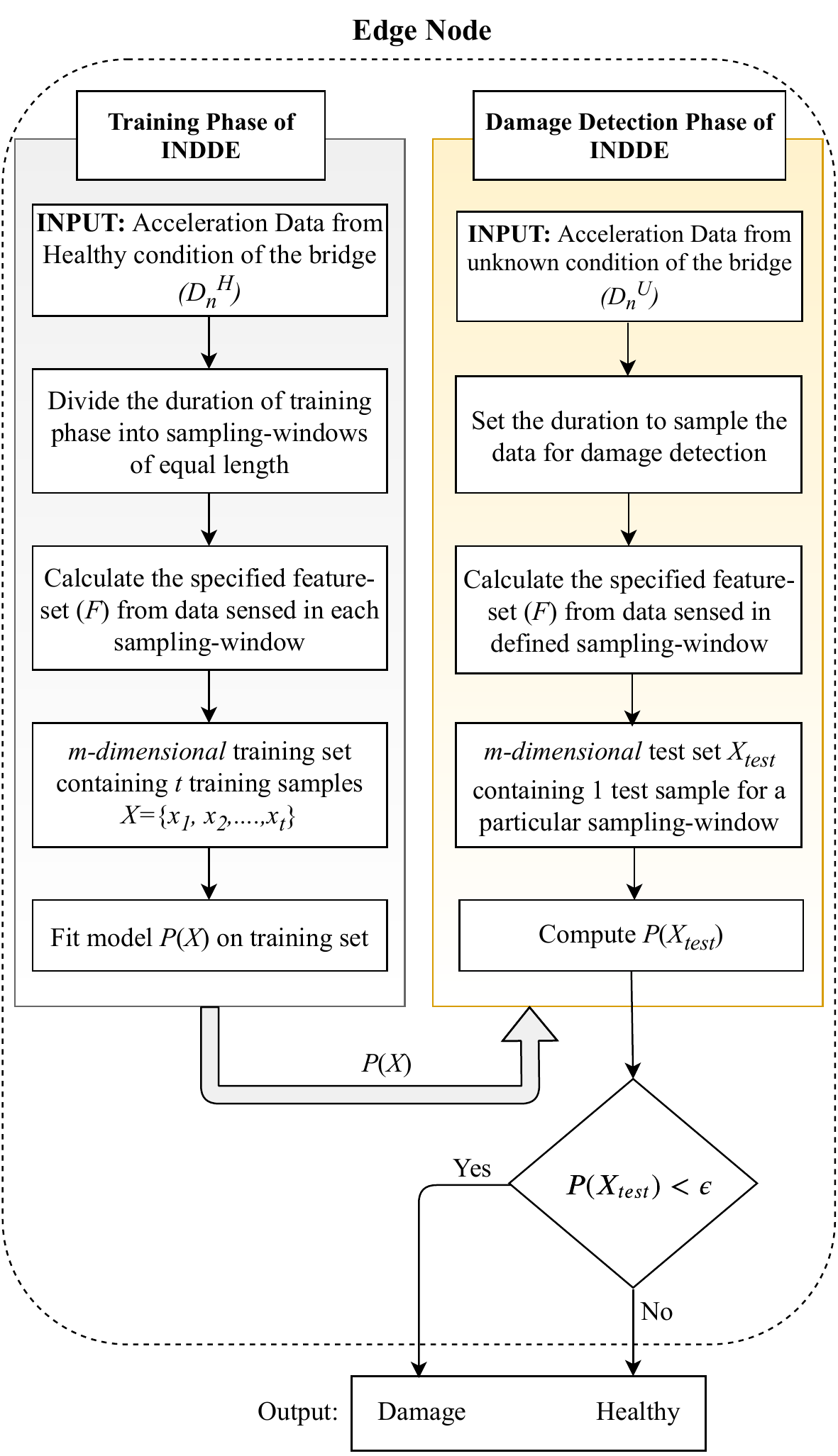}
\caption{Workflow of INDDE}
\label{mechanism}
\end{figure}

We consider a set of $N$ sensor nodes mounted on different locations of a bridge to sense, monitor, and transmit the information concerning to the structural event. Each sensor node can perform computation on sensed acceleration data to detect the damage in their vicinity. The data points collected concerning to the vibration responses are assumed to be a vector as $D_n=[d_1, d_2, d_3,.....,d_k];$ where $n=1,....,N$ are the sensor nodes and $k$ is the total number of data points sensed by a node $n$ over a time duration. For the sake of simplicity, the set of data points corresponding to the ``healthy" condition or reference state of the bridge will be denoted as $D^H$ and that corresponding to the subsequent unknown condition as $D^U$. The acceleration data collected when bridge is healthy, i.e., $D^H$, cover various environmental and ambient conditions as well as operational conditions, such as traffic loading. Figure \ref{mechanism} illustrates the mechanism of INDDE running at each sensor node and considers $D^H$ as reference data for damage detection. The functioning of INDDE is broadly divided into two phases: training phase and damage detection phase. In the training phase, INDDE fits a model by extracting the statistical features from raw acceleration data in healthy condition of the bridge. The trained model is used to further classify the raw acceleration measurement from unknown condition of the bridge as either healthy or damaged.

\subsection{Phase I: Train the model}
\label{train}
\begin{figure*}[ht!]
\centering
\includegraphics[scale=0.9]{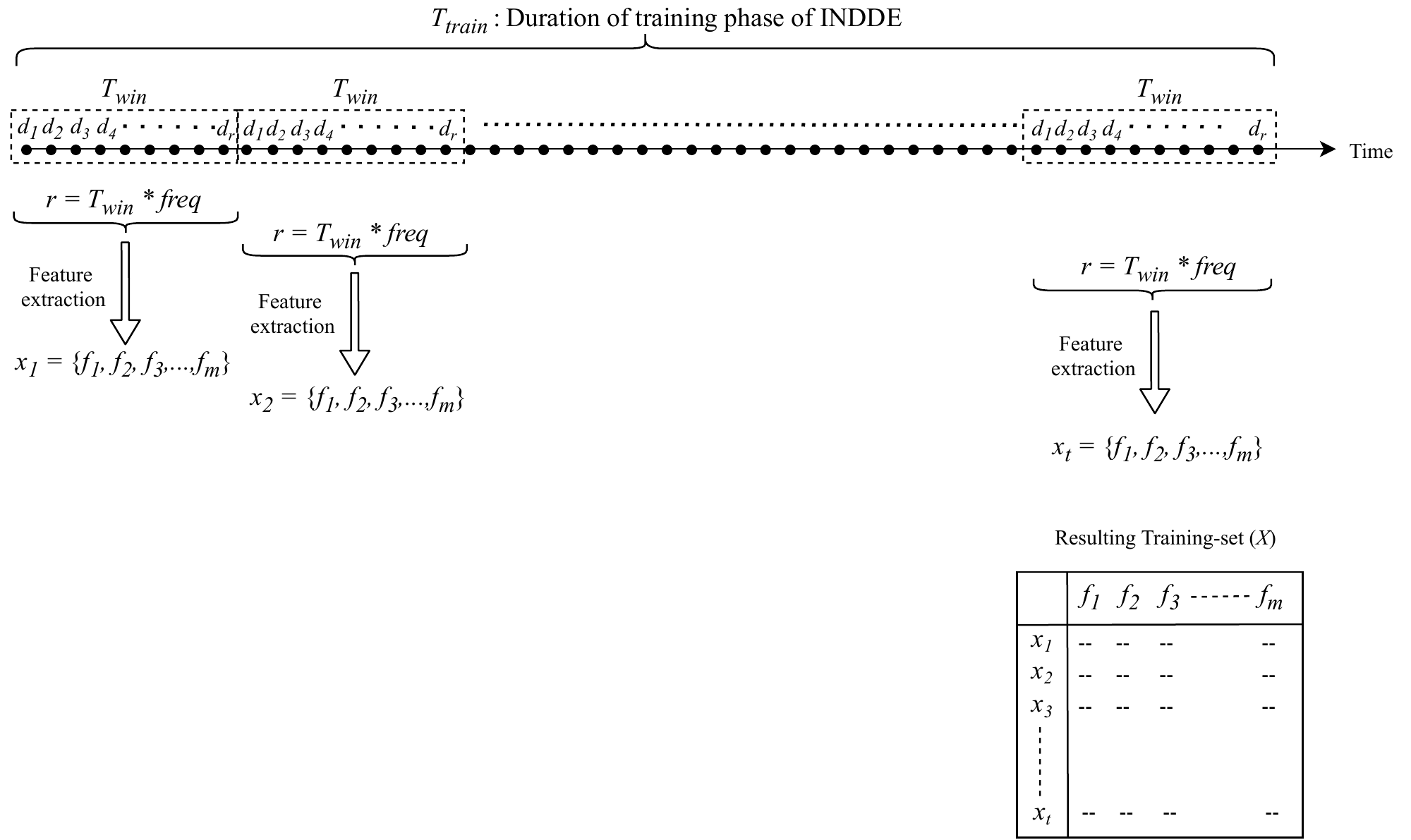}
\caption{Illustration of training phase of INDDE}
\label{dd_working}
\end{figure*}

A training-set $X = \{x_1, x_2, x_3,.....x_t\}$ is created from raw acceleration measurements corresponding to healthy condition of the bridge. Each training sample $x_i \in X$ is an $m-dimensional$ feature vector $F=\{f_1, f_2, f_3,.....,f_m\}$, where $f_1, f_2, f_3,....., f_m$ are the statistical features extracted from sensed acceleration data in healthy condition of the bridge, such that $f_1 \sim \mathcal{N}(\mu_1, \sigma_1^2)$, $f_2 \sim \mathcal{N}(\mu_2, \sigma_2^2)$, $f_3 \sim \mathcal{N}(\mu_3, \sigma_3^2)$,......, $f_m \sim \mathcal{N}(\mu_m, \sigma_m^2)$ are normally distributed with their respective mean ($\mu$) and standard deviation ($\sigma$). The training-set $X$ is a multivariate dataset consisting $t$ observations, which can be represented as $t$ points in a $m-dimensional$ space. The total number of tuples ($t$) in $X$ depends on the duration of sampling-window or the number of data points in a sampling window for calculating the features. For instance, the training period ($T_{train}$) for a sensor node, measuring acceleration data in healthy condition of bridge at sampling frequency $freq = 100Hz$, is set to 6 hours, i.e., $k=21,60,000$ data points will be recorded in $D^H$. If a time-window of 5 minutes ($T_{win}=300$ seconds) is set to calculate the feature vector from sensed data points ($r = T_{win} \times freq = 30,000$ data points), the total number of training samples in $X$ will be $k/r$, i.e., $t = 72$. Likewise, if duration of time-window is set as 10 minutes, it will result in 36 training samples in $X$. Figure \ref{dd_working} depicts the underlying steps to create training-set ($X$) in the training phase of INDDE.

The time-domain analysis of structural vibration responses allows extracting signal parameters in a simpler way than the modal analysis \cite{farrar}. Keeping this in mind, seven statistical parameters, i.e, mean, mean square, standard deviation, variance, skewness, kurtosis, and crest factor, are considered in this paper because a single statistic is not enough to detect the damage. The condition of the bridge can be characterized by these statistical features which allow working directly with acceleration measurements in the time domain. Each parameter depict unique characteristic of the vibration signal and the values of these parameters varies with the condition of the bridge, thus can be used for damage detection.

Mean and mean square are the measures of central tendency of the data. Standard deviation and variance are the measures of variability which represents the dispersion related to mean value. Skewness represents symmetry of data distribution and kurtosis represents the flattening of the distribution considering Gaussian curve. Crest factor indicates how extreme the peaks are in a signal. Although some statistics are correlated, all of them are used as inputs for INDDE, without employing any feature selection method beforehand. The mentioned statistical features can be calculated as follows from the data-points observed in a sampling-window. 

\begin{equation}
f_1:\ mean\ (\overline{m}) =  \frac{1}{r}\sum_{i=1}^r d_i
\end{equation}

\begin{equation}
f_2:\ mean\ square\ (m_{sq}) = \frac{1}{r} \sum_{i=1}^r (d_i)^2
\end{equation}

\begin{equation}
f_3:\ variance\ (\sigma^2) = \frac{1}{r} \sum_{i=1}^r (d_i - \overline{m})^2
\end{equation}

\begin{equation}
f_4:\ standard\ deviation\ (\sigma) = \sqrt{\frac{1}{r} \sum_{i=1}^r (d_i - \overline{m})^2}
\end{equation}

\begin{equation}
f_5:\ skewness\ (skw) =  \frac{1}{r} \sum_{i=1}^r (d_i - \overline{m})^3  {\sigma^3}
\end{equation}

\begin{equation}
f_6:\ kurtosis\ (krt) = \frac{1}{r} \sum_{i=1}^r (d_i - \overline{m})^4 {\sigma^4}
\end{equation}

\begin{equation}
f_7:\ Crest\ factor\ (cf) = max(d){\sqrt{m_{sq}}}
\end{equation}

The training-set $X$ is used to estimate the probability density function (PDF) corresponding to the healthy condition of bridge. We used multivariate Gaussian kernel function to estimate the PDF with parameters $\Omega$ and $\Sigma$, where $\Omega \in \mathbb{R}^m$ is a $m \times 1-dimensional$ vector of means corresponding to all features, i.e., $\Omega = [E(f_1), E(f_2), E(f_3),...., E(F_m)]$, and $\Sigma \in \mathbb{R}^{m \times m}$ is a $m \times m$ covariance matrix, identified from $X$. The covariance for a pair of components $u$ and $v$ is calculated by Eq. \ref{cov}, which can further be used to create $m \times m$ covariance matrix as depicted in Eq. \ref{covm}.

\begin{equation}
\sigma_{uv}=E[x_u x_v]-E[x_u]E[x_v]
\label{cov}
\end{equation}

\begin{equation}
\Sigma=
\begin{bmatrix} 
\sigma_{11} & \sigma_{12} & \sigma_{13} &.... &\sigma_{1m} \\
\sigma_{21} & \sigma_{22} & \sigma_{23} &.... &\sigma_{2m} \\
.   &.  &.  &.  &.  \\
.   &.  &.  &.  &.  \\
\sigma_{m1} & \sigma_{m2} & \sigma_{m3} &.... &\sigma_{mm}
\end{bmatrix}
\label{covm}
\end{equation}

The PDF of multivariate Gaussian distribution for training set $X$ is given by Eq. \ref{pdf}.

\begin{equation}
\label{pdf}
P(X; \Omega, \Sigma) = \frac{1}{(2 \pi)^{m/2} |\Sigma|^{1/2}} exp\Big(-\frac{1}{2} (X - \Omega)^T \Sigma^{-1} (X-\Omega) \Big)
\end{equation}

Where, $|\Sigma|$ is the determinant of matrix $\Sigma$. The estimated PDF is used in damage detection phase of INDDE. The condition of the bridge is identified as healthy or damaged on the basis of the PDF magnitude corresponding to the test-data from unknown condition of the bridge. 

\subsection{Phase II: Damage detection}
\label{detect}
\begin{figure*}[ht!]
\centering
\includegraphics[scale=0.9]{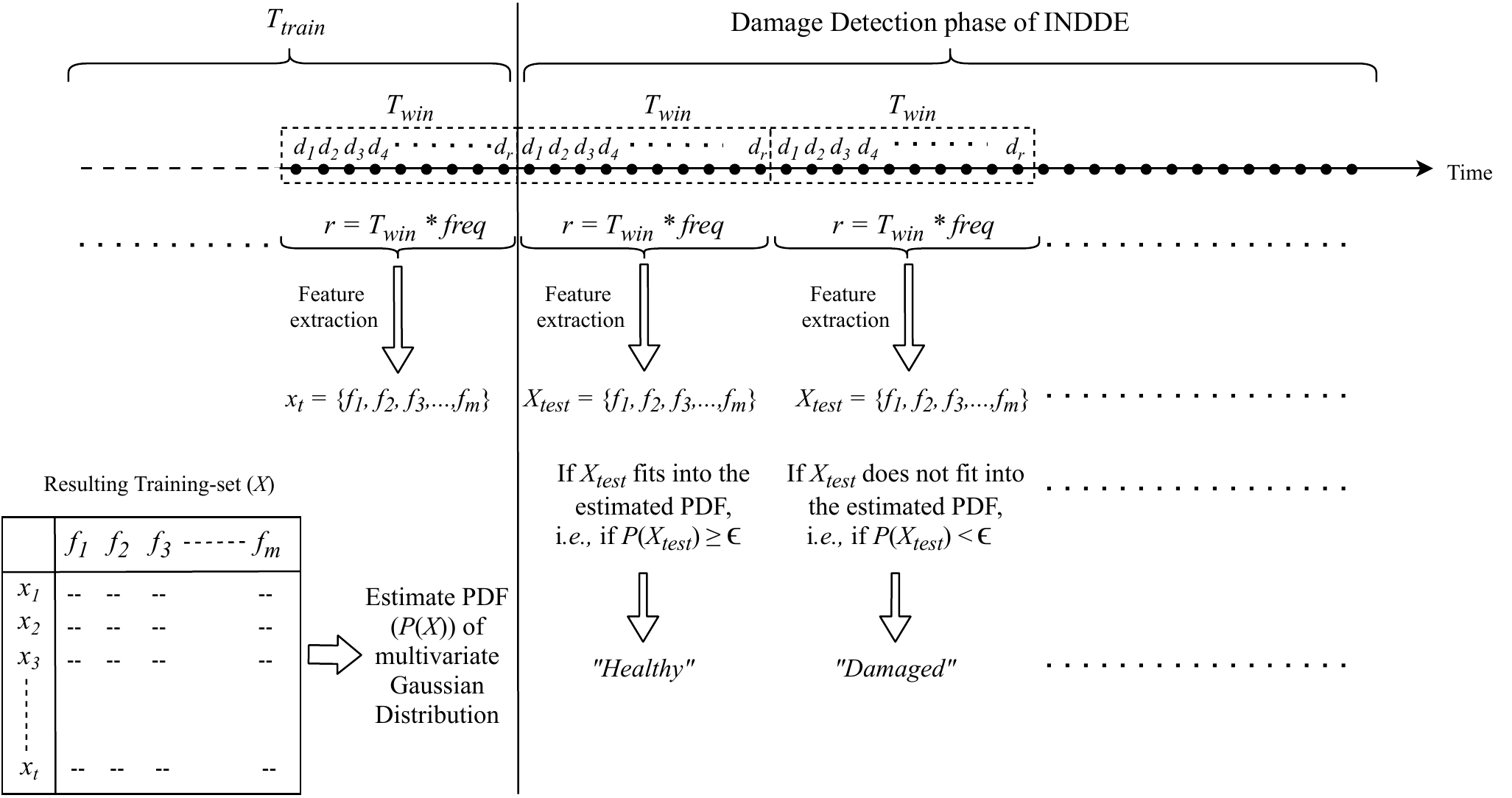}
\caption{Illustration of damage detection phase of INDDE}
\label{detection}
\end{figure*}

The idea behind damage detection mechanism of INDDE is to extract the same statistical feature set ($F$) from acceleration data corresponding to the unknown condition of the bridge in order to estimate whether it comes from the same distribution as the training-set or not. Figure \ref{detection} illustrates the damage detection phase of INDDE. 

A sampling-window of same time-duration ($t_{win}$) is set to perform the damage detection at a regular time-interval after the completion of training phase. The raw acceleration measurements sensed over the time-window are represented as $D^U=\{d_1, d_2, d_3,......, d_k \}$ which corresponds to the unknown condition of the bridge. These sensed data points are transformed into $1 \times m - dimensional$ test dataset $X_{test}$ consisting statistical features that were used in the training phase. The health of the bridge is characterized by estimating the probability corresponding to $X_{test}$ using Eq. \ref{pdf}.

\begin{center}
\[    Condition =
\begin{cases}
    Damaged,				& \text{if } P(X_{test}; \Omega, \Sigma) < \epsilon \\
	Healthy,              & \text{if } P(X_{test}; \Omega, \Sigma) \geq \epsilon 
\end{cases}
\]
\end{center}

where, $\epsilon$ is a threshold parameter that can be set by cross-validating the trained model to decide whether new observation data represents damage or not.

\subsection{Computational complexity analysis of INDDE}
In order to analyse the computation complexity of INDDE, it is required to analyse the complexities of various steps involved in INDDE, such as creating training set $X$, calculating $\Omega$, $\Sigma$, $|\Sigma|$, and $\Sigma^{-1}$ for PDF estimation.

Creating training set $X$ is an iterative process over a long period of time divided into $t$ time-windows. Since, all the seven statistical parameters are identified over $r$ collected data points in each time-window, the creation of training set is done in $O(tr)$ time. Calculating $\Omega$, $\Sigma$, $|\Sigma|$, and $\Sigma^{-1}$ on the training-set takes $O(tm)$, $O(m^2)$, $O(m^3)$, and $O(m^2)$ time respectively. All the arithmetic operations take constant time, i.e., $O(1)$. Thus, the time complexity of the whole process of INDDE is $[O(m^3)+O(tm)+O(tr)]$ after ignoring the low-order terms.

\section{Case Studies}
\label{casestudy}
In this section, two comprehensive case studies are investigated to illustrate the efficiency of the proposed damage detection mechanism. The first case study uses the real data-set obtained from Yonghe Bridge, and the second case study uses acceleration data generated from a lab-scale setup of a steel beam subjected to progressive damages at multiple locations.

\subsection{Case Study on Yonghe Bridge}
\label{case}
\begin{figure*}
\centering
\includegraphics[scale=0.8]{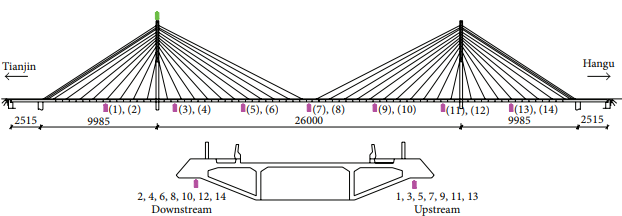}
\caption{Yonghe bridge elevation \cite{kaloop}}
\label{bridge}
\end{figure*}

The Yonghe bridge is a cable-stayed bridge constructed in Mainland China connecting Tianjin and Hangu cities \cite{kaloop}\cite{kaloop2}. Harbin Institute of Technology Research Center has established the SHM system to sense, monitor, and transmit the acceleration data from fourteen uniaxial accelerometers mounted along the deck of the main span of the bridge (see Figure \ref{bridge}). The sampling frequency of the sensors is 100 Hz. The acceleration data in healthy and damaged condition of the bridge were recorded by 14 deck sensors for 24 hours on January 17, 2008 and July 31, 2008 respectively. Figure \ref{healthy} and \ref{damaged} show the acceleration data acquired by node 1 in one hour corresponding to the healthy and damaged condition of the bridge respectively. The variation in the acceleration is presumably due to the fact that as the bridge is damaged, it needs to resist external loads using more kinetic energy as it attains less potential energy. The variations in the acceleration data can be characterized by statistical features concerning to state of the bridge and can be used to detect the damage, if occurs.

\subsubsection{Experimental Data}
\begin{figure}[ht!]
\centering
	\begin{subfigure}{0.48\textwidth}
 	\includegraphics[width = \textwidth]{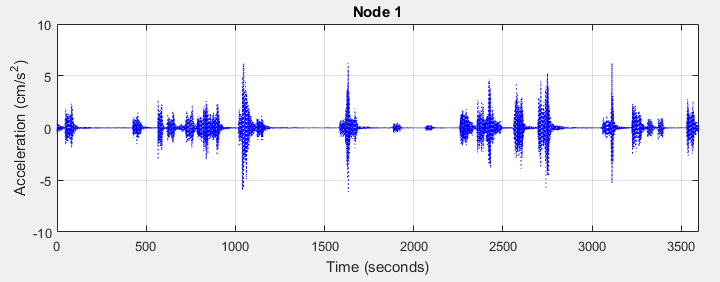}
 	\caption{Vibration responses in healthy condition of bridge}
 	\label{healthy}
	\end{subfigure}%
	\hspace{0.3cm}
	\begin{subfigure}{0.48\textwidth}
	\includegraphics[width = \textwidth]{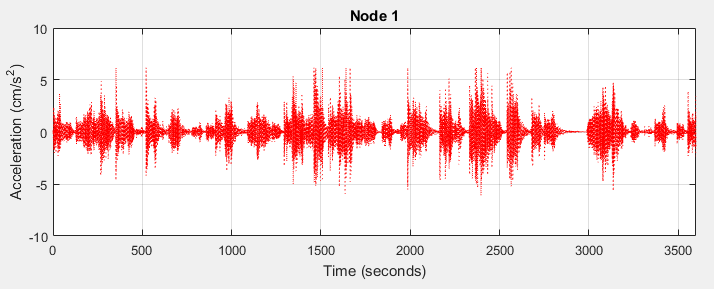}
	\caption{Vibration responses in damaged condition of bridge}
	\label{damaged}
	\end{subfigure}
\caption{Vibration responses of one hour duration acquired by sensor node (1) deployed at the bridge}
\label{data}
\end{figure}

We have raw acceleration data from both healthy and damaged condition of the Yonghe bridge. Figure \ref{data} shows raw acceleration data for one hour corresponding to the healthy and damage condition of the bridge. Since INDDE is an on-node real-time damage detection mechanism, the data from healthy condition is used in the training phase to emulate the real-world scenario. After completing the training, damaged data is used to test the damage detection accuracy of INDDE.

The time duration of training phase is set to 18 hours (i.e., 64,80,000 data-points) for each sensor node to extract the selected statistical features and build the model. For that duration, a time-window $T_{win}$ of 5 minutes is fixed to create a tuple in the training set, i.e.,  each sensor node extract the statistical features from $r = 30,000$ ($ t_{win} \times freq$) raw acceleration data-points collected in that duration. It results in a training set $X_{216 \times 7}$ containing $t = 216$ ($=64,80,000/30,000$) samples in 7-dimensional space to estimate the PDF of multivariate Gaussian distribution using Eq. \ref{pdf}. Since the condition of bridge is already known to be healthy, the estimated PDF can be used to detect any abnormality or damage from newly sensed data points, as discussed in Section \ref{detect}. The remaining data of 6 hours (i.e., 21,60,000 data points, equivalent to 72 test samples) corresponding to healthy condition are used to cross-validate the model. In addition, the data of 24 hours duration (i.e., 86,40,000 data points, equivalent to 288 test samples) corresponding to damaged bridge condition is used for testing and evaluating the damage detection accuracy of INDDE.

\begin{figure*}[ht!]
	\begin{subfigure}{0.5\textwidth}
 	\includegraphics[width = \textwidth]{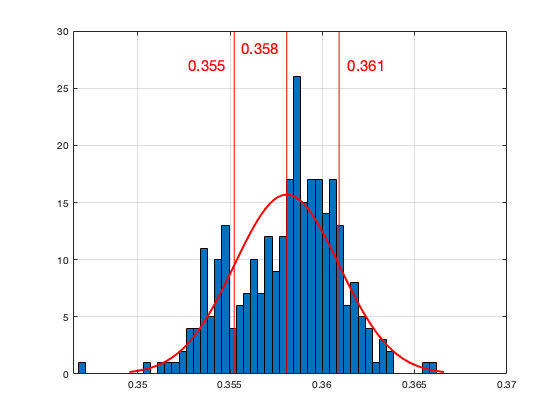}
 	\caption{Statistical feature $f_1$}
 	\label{f1}
	\end{subfigure}%
	\hspace{0.2cm}
	\begin{subfigure}{0.5\textwidth}
	\includegraphics[width = \textwidth]{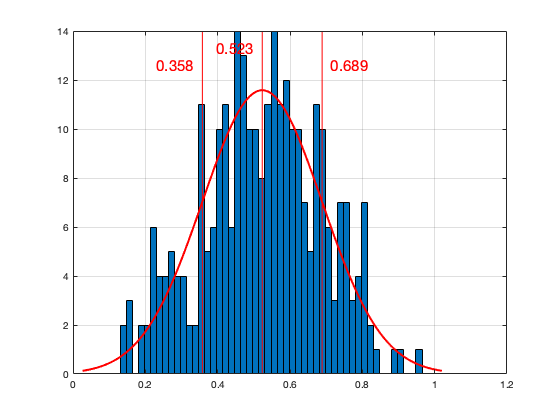}
	\caption{Statistical feature $f_2$}
	\label{f2}
	\end{subfigure}
	\\ \vspace{0.6cm}
	\begin{subfigure}{0.5\textwidth}
 	\includegraphics[width = \textwidth]{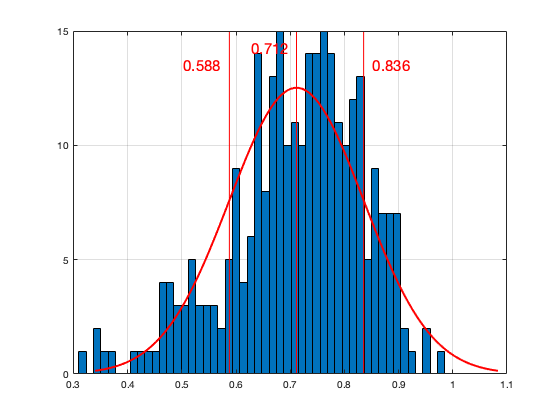}
 	\caption{Statistical feature $f_3$}
 	\label{f3}
	\end{subfigure}%
	\hspace{0.2cm}
	\begin{subfigure}{0.5\textwidth}
	\includegraphics[width = \textwidth]{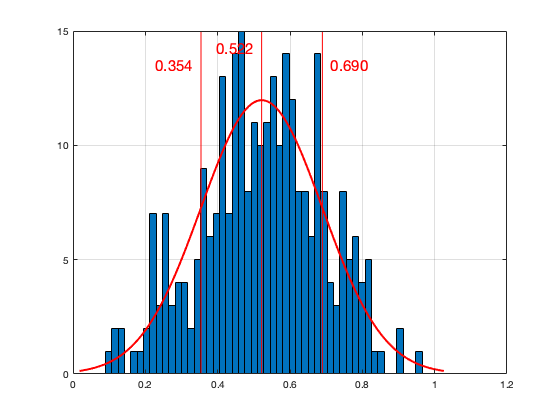}
	\caption{Statistical feature $f_4$}
	\label{f4}
	\end{subfigure}
	\\ \vspace{0.6cm}
	\begin{subfigure}{0.5\textwidth}
 	\includegraphics[width = \textwidth]{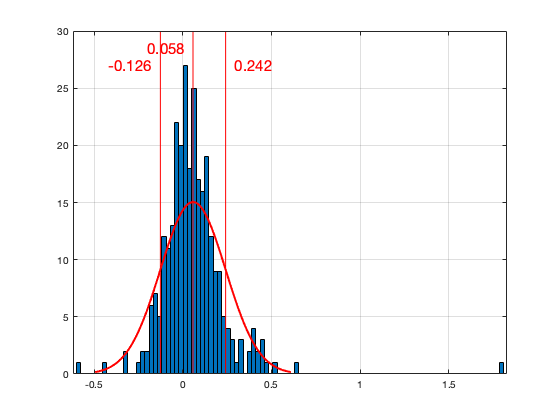}
 	\caption{Statistical feature $f_5$}
 	\label{f5}
	\end{subfigure}%
	\hspace{0.2cm}
	\begin{subfigure}{0.5\textwidth}
	\includegraphics[width = \textwidth]{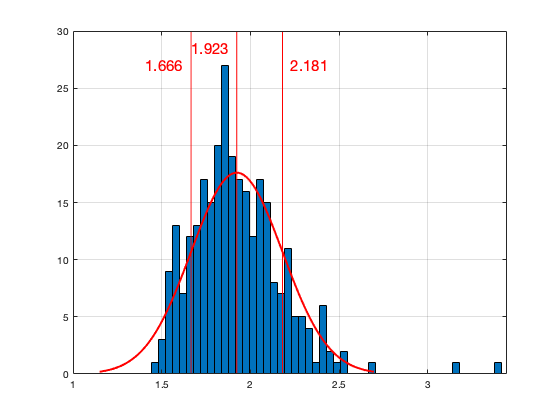}
	\caption{Statistical feature $f_6$}
	\label{f6}
	\end{subfigure}
\end{figure*}
\begin{figure}\ContinuedFloat
	\centering
	\begin{subfigure}{0.5\textwidth}
	\includegraphics[width = \textwidth]{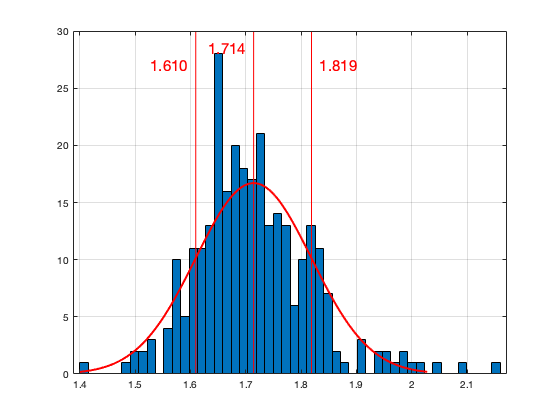}
	\caption{Statistical feature $f_7$}
	\label{f7}
	\end{subfigure}
\caption{Histogram of statistical features with Gaussian distribution fit corresponding to training samples at sensor node (1)}
\label{normalplot}
\end{figure}

The reason behind choosing Gaussian kernel function to fit the training-set is the characteristics of the statistical features.  Figure \ref{normalplot} depicts the histogram of statistical features $f_1$ to $f_7$ to assess the distribution of training data at sensor node (1). The red curve superimposed on the histograms represents bell-shaped ``normal" curve with a mean and mean+/-standard deviation represented by red vertical lines. The multivariate Gaussian distribution is useful in analyzing the relationship between multiple normally distributed variables, i.e., statistical features here, which are also found to be jointly normal. Therefore, it leads to estimate the PDF of multivariate Gaussian distribution, as mentioned in Section \ref{train}, from acceleration data observed in healthy condition of the bridge.

\subsubsection{Performance analysis of INDDE}
\label{result}
The acceleration data from six different sensor nodes (namely (1), (2), (6), (7), (13), and (14)) located at different locations along the span of the bridge (see Figure \ref{bridge}) are selected to evaluate the proposed damage detection mechanism. Each sensor node execute training phase of INDDE and estimate PDF individually corresponding to their training-set. 216 training samples generated from 18 hours of acceleration data from healthy condition were used for the training phase, whereas the remaining 6 hours of healthy data (i.e., 72 samples) and 24 hours of damage data (i.e., 288 samples) were used for the testing of INDDE. Damage detection results from test data are represented in terms of the number of true positives (TP) (i.e., healthy condition is correctly identified), true negatives (TN) (i.e., damaged condition is correctly identified), false positives (FP) (i.e., damaged condition is identified as healthy), and false negatives (FN) (i.e., healthy condition is identified as damaged) for all the sensors mentioned above (see Figure \ref{confusion}).

\begin{figure}[h!]
\centering
\renewcommand{\arraystretch}{1.2}
\begin{tabular}{llcc}
                                             &                              & \multicolumn{2}{c}{Detected}                                \\ \cline{3-4} 
                                             & \multicolumn{1}{l|}{}        & \multicolumn{1}{c|}{Healthy} & \multicolumn{1}{c|}{Damaged} \\ \cline{2-4} 
\multicolumn{1}{c|}{\multirow{2}{*}{Actual}} & \multicolumn{1}{c|}{Healthy} & \multicolumn{1}{c|}{TP}      & \multicolumn{1}{c|}{FP}      \\ \cline{2-4} 
\multicolumn{1}{c|}{}                        & \multicolumn{1}{c|}{Damaged} & \multicolumn{1}{c|}{FN}      & \multicolumn{1}{c|}{TN}      \\ \cline{2-4} 
\end{tabular}
\caption{Confusion matrix}
\label{confusion}
\end{figure}

The confusion matrices for aforementioned six sensors are represented in tabular form in TABLE \ref{conf}. These parameters are used to evaluate INDDE on different performance measures, such as accuracy, precision, recall, F-score, type-I error (false positive error), and type-II error (false negative error) \cite{incontextiot}, as shown in TABLE \ref{comp}.

\begin{table}[ht!]
\centering
\caption{The number of TP, TN, FP, and FN for different sensor nodes deployed on the bridge}
\label{conf}
\renewcommand{\arraystretch}{1.2}
\begin{tabular}{|p{1cm}|p{1cm}|p{1cm}|p{1cm}|p{1cm}|}
\hline
Nodes		&TP		&TN	&FP		&FN \\ \hline
1     &72      &275    &13   &0     \\ \hline
2     &71      &288    &1     &0     \\ \hline
6     &72     &288    &0     &0     \\ \hline
7     &71      &288    &0     &1		 \\ \hline
13   &72     &288     &0     &0      \\ \hline
14    &72      &288    &0     &0      \\ \hline
\end{tabular}
\end{table}

Although no data from damaged event has been employed to construct the model and data obtained only from healthy state of the bridge have been used for the training, the trained model can successfully identify the damages with $\approx 96-100\%$ of accuracy, as can be seen in Table \ref{comp}.

\begin{table}[ht!]
\centering
\caption{Performance evaluation of INDDE across different sensor nodes}
\label{comp}
\renewcommand{\arraystretch}{1.2}
\begin{tabular}{|p{1cm}|p{1.5cm}|p{1.2cm}|p{1.2cm}|p{1.2cm}|p{1.2cm}|p{1.2cm}|}
\hline
Nodes & Accuracy (in \%) & Precision & Recall & F-score & Type-I error & Type-II error \\ \hline
1     &96.39\%    &0.84    &1   &0.91   &0.045   &0      \\ \hline
2     &99.7\%      &1       &0.98   &0.99       &0      &0.01    \\ \hline
6     &100\%    &1       &1      &1        &0         &0               \\ \hline
7     & 99.7\%      &1      &0.98     &0.99        &0    &0.013  \\ \hline
13    &100\%     &1     &1        &1     &0         &0               \\ \hline
14    &100\%      &1    &1     &1         &1          &1               \\ \hline
\end{tabular}
\end{table}

To further investigate the efficiency of the proposed methodology, a separate approach was adopted as follows. We considered a total of 432 samples (216 samples each) from healthy and damaged condition of the bridge to train different supervised learning models, such as decision tree (DT), random forest (RF), neural network (NN), and support vector machine (SVM). All these models are tested with a total of 144 test samples (72 samples each) from both healthy and damaged scenario. The models are trained and tested for sensors 1, 2, 6, 7, 13, and 14. A comparison of these methods on the basis of overall accuracy of correctly classifying the states (healthy or damaged) of the bridge is depicted in TABLE \ref{compmodel}.

Although INDDE does not require data from damaged bridge condition to train the model and only data from healthy condition have been used, it achieves accuracy as good as supervised learning models. Furthermore, INDDE performs damage detection in real-time by the raw acceleration on the node itself. Therefore, it is advantageous for the resource-constrained sensor nodes due to its low computation cost and no message exchange policy in the damage detection process.

\begin{table}[ht!]
\centering
\caption{Comparison of accuracy of different models for correctly classifying the states of the bridge}
\label{compmodel}
\renewcommand{\arraystretch}{1.2}
\begin{tabular}{|p{1cm}|p{1.2cm}|p{1.2cm}|p{1.2cm}|p{1.2cm}|p{1.2cm}|}
\hline
Nodes & DT & RF & SVM & NN & INDDE  \\ \hline
1     &100\%    &99.3\%    &98.6\%   &88.8\%   &96.39\%  \\ \hline
2     &100\%    &100\%     &98.6\%   &99.3\%   &99.7\%   \\ \hline
6     &100\%    &100\%      &99.3\%  &99.3\%   &100\%   \\ \hline
7     &100\%     &100\%     &99.3\%  &100\%     &99.7\%  \\ \hline
13    &100\%    &100\%     &99.3\%  &84.7\%    &100\%   \\ \hline
14    &100\%    &100\%     &99.3\%   &100\%     &100\%   \\ \hline
\end{tabular}
\end{table}

\subsection{Case Study on Steel Beam}
\begin{figure*}[ht!]
\centering
\includegraphics[scale=0.1]{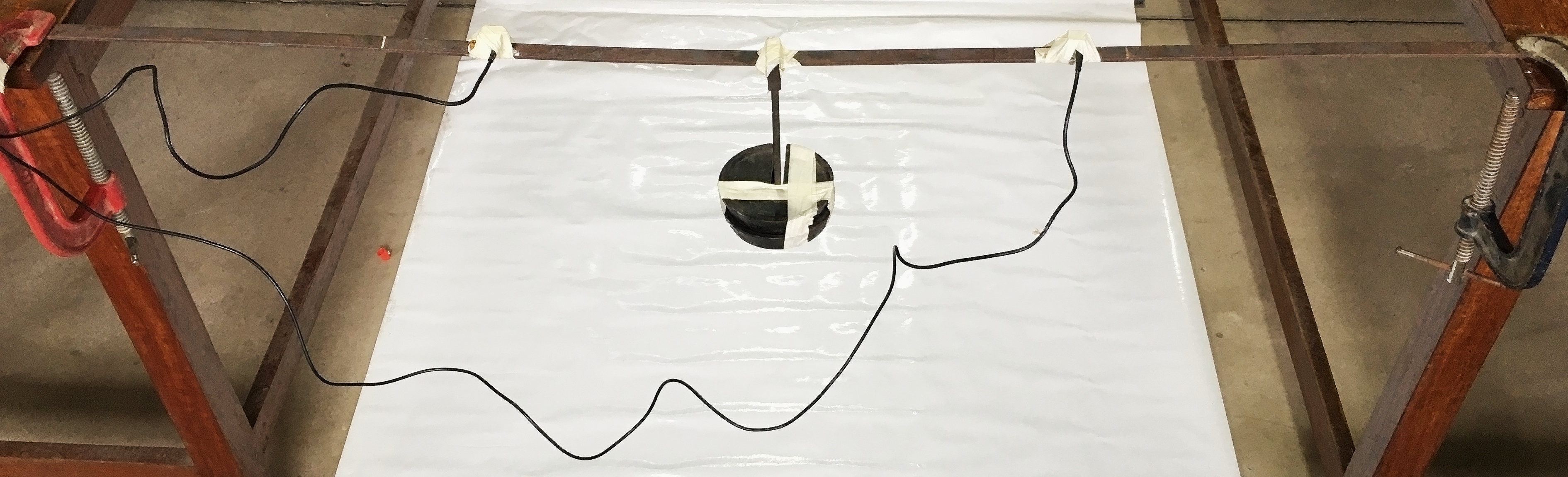}
\caption{Fixed-end cantilever beam subjected to a central weight of 2 KGs.}
\label{testbed}
\end{figure*}

\begin{figure*}[ht!]
\centering
\includegraphics[scale=0.42]{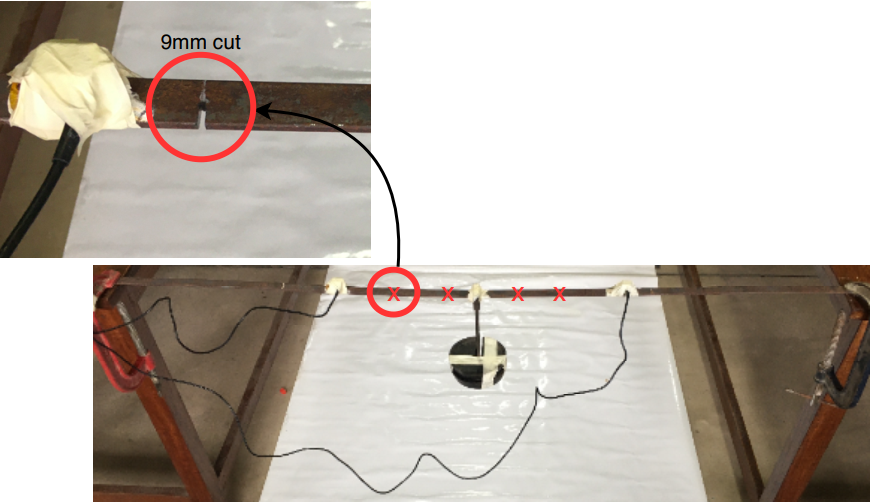}
\caption{The notches introduced in beam at different locations to induce damage}
\label{damage}
\end{figure*}

\begin{figure*}[ht!]
\centering
\includegraphics[scale=0.45]{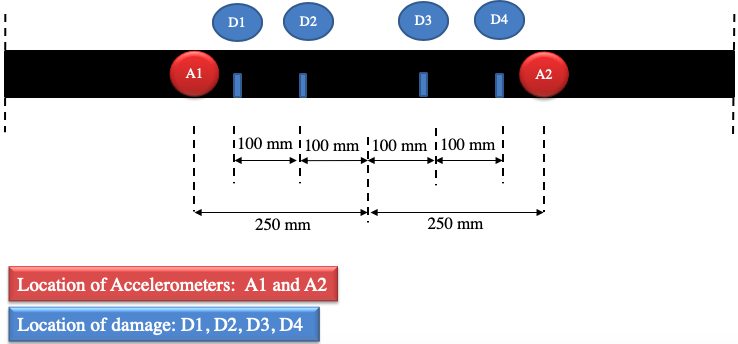}
\caption{An illustration of test setup and damage scenarios}
\label{illustrate}
\end{figure*}

In this case study, an experimental test-bed was established to further verify the capability of INDDE to detect the damage in a steel beam structure. The test setup was comprised of a steel beam with fixed support at its both ends to emulate a simplistic bridge structure (see Figure \ref{testbed}). The length, width, and thickness of the beam were 1,240 mm, 18 mm, and 4 mm respectively. The mass of steel beam used was 661.6 grams and a weight of 2 KGs was hung at the center of beam to excite the beam under free vibrations. Two accelerometers (ARJ-200A) were evenly placed on the beam, i.e., at 250 mm each side from the center, to measure the vertical acceleration of the beam with a sampling frequency of 200 Hz. The vibration measurements were recorded by connecting the 	accelerometers to a four channel data logger (TML DC-204Ra dynamic strain recorder). The vibration measurements were collected for a duration of 8 minutes in undamaged scenario. After acquiring the acceleration data from intact steel beam, damages were progressively introduced in the structure by cutting the notches of different depth at different locations on the beam and measurements were recorded in each damage scenario. Figure \ref{damage} and \ref{illustrate} show the beam that was subjected to different damages over a course of 48 minutes.

\subsubsection{Experimental Data}
The free vibration data in case of undamaged (healthy) scenario and each damaged scenario of the beam were observed for 8 minutes and 6 minutes, respectively. The beam was subjected to eight different damages by cutting the notches of different depths, thus results in 48 minutes of damaged data. After acquiring the data-set, we used INDDE to build the model corresponding to undamaged state of the beam and validate the trained model by using the data acquired from damage scenarios, similar to previous case study in Section \ref{case}.

\begin{figure*}[ht!]
	\begin{subfigure}{0.5\textwidth}
 	\includegraphics[width = \textwidth]{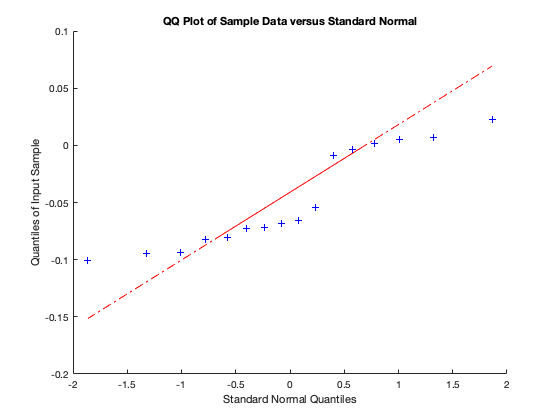}
 	\caption{Statistical feature $f_1$}
 	\label{f1}
	\end{subfigure}%
	\hspace{0.2cm}
	\begin{subfigure}{0.5\textwidth}
	\includegraphics[width = \textwidth]{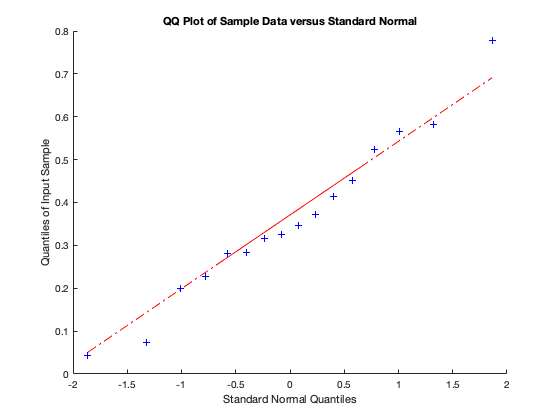}
	\caption{Statistical feature $f_2$}
	\label{f2}
	\end{subfigure}
	\\ \vspace{0.3cm}
	\begin{subfigure}{0.5\textwidth}
 	\includegraphics[width = \textwidth]{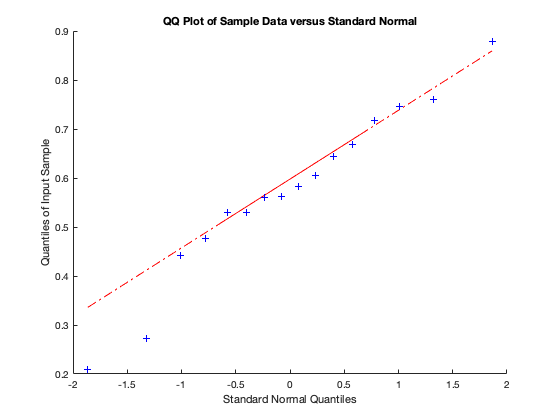}
 	\caption{Statistical feature $f_3$}
 	\label{f3}
	\end{subfigure}%
	\hspace{0.2cm}
	\begin{subfigure}{0.5\textwidth}
	\includegraphics[width = \textwidth]{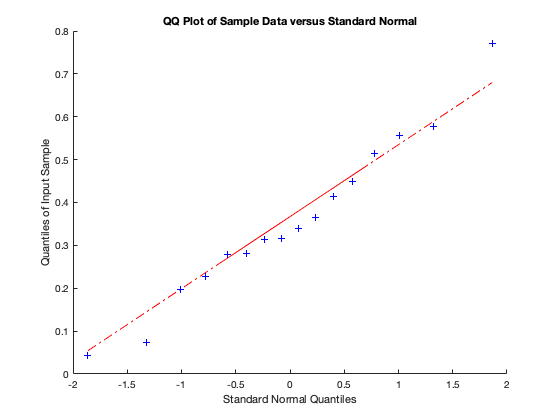}
	\caption{Statistical feature $f_4$}
	\label{f4}
	\end{subfigure}
	\\ \vspace{0.3cm}
	\begin{subfigure}{0.5\textwidth}
 	\includegraphics[width = \textwidth]{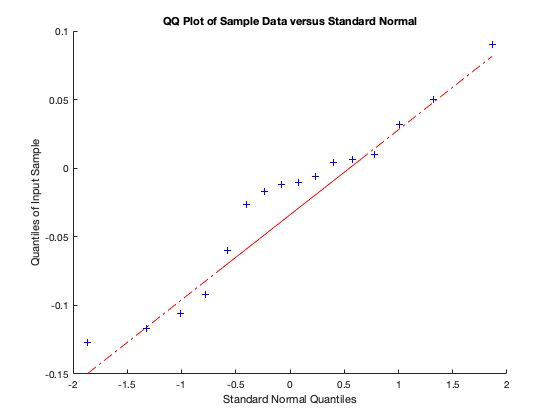}
 	\caption{Statistical feature $f_5$}
 	\label{f5}
	\end{subfigure}%
	\hspace{0.2cm}
	\begin{subfigure}{0.5\textwidth}
	\includegraphics[width = \textwidth]{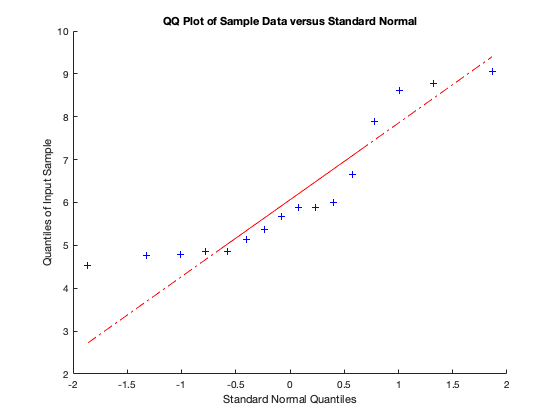}
	\caption{Statistical feature $f_6$}
	\label{f6}
	\end{subfigure}
\end{figure*}
\begin{figure}\ContinuedFloat
\centering
	\begin{subfigure}{0.5\textwidth}
	\includegraphics[width = \textwidth]{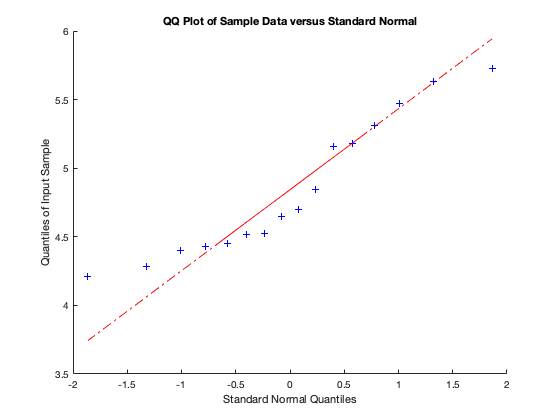}
	\caption{Statistical feature $f_7$}
	\label{f7}
	\end{subfigure}
\caption{Normal Q-Q plot of statistical features corresponding to training samples}
\label{qqplot}
\end{figure}

The time duration of training phase ($T_{train}$) is set to 6 minutes (i.e., 48,000 data points) for both the sensor nodes. A time window ($T_{win}$) of 30 seconds (i.e., $r=6000$ data points) is fixed to extract the statistical features from the sensed data points to create a tuple in the training set. The resulted training set $X_{12 \times 7}$ contains $t=12$ samples in 7-dimensional space to estimate the PDF of multivariate Gaussian distribution by using Eq. \ref{pdf}. The remaining data points of 2 minutes (equivalent to 4 test samples) corresponding to healthy condition of the beam are used to cross-validate the model. In addition, the data of 48 minutes (i.e., 96 test samples) corresponding to damaged scenarios were used for testing and evaluating the damage detection accuracy of INDDE. The distribution of statistical features $f_1$ to $f_7$ from training set of node 1 against the expected normal distribution are demonstrated by Q-Q plots (Quantile-Quantile plots) in Figure \ref{qqplot}. 

\subsubsection{Performance analysis of INDDE}
The training phase is executed on the training samples of both the sensor nodes to estimate the PDF individually. 12 training samples generated from 6 minutes of acceleration data from healthy condition were used to estimate the PDF, whereas the remaining 2 minutes of healthy data (i.e., 4 samples) and 48 minutes of damage data (i.e., 96 samples) were used for the testing of INDDE in this lab-scale use-case scenario. The confusion matrix corresponding to both the sensor nodes are shown in Figure \ref{cf1} and \ref{cf2}, respectively. Based on these confusion matrices, different performance measures, such as accuracy, precision, recall, F-score, type-I error, and type-II error are identified and shown in TABLE \ref{perf}. The accuracy of the trained model depends on the number of training samples. Since the number of training samples in this case study is very less as compared to the previous case study, it results in low accuracy in damage detection process.
\begin{figure}[ht!]
\centering
\renewcommand{\arraystretch}{1.2}
\begin{tabular}{llcc}
                                             &                              & \multicolumn{2}{c}{Detected}                                \\ \cline{3-4} 
                                             & \multicolumn{1}{l|}{}        & \multicolumn{1}{c|}{Healthy} & \multicolumn{1}{c|}{Damaged} \\ \cline{2-4} 
\multicolumn{1}{c|}{\multirow{2}{*}{Actual}} & \multicolumn{1}{c|}{Healthy} & \multicolumn{1}{c|}{4}      & \multicolumn{1}{c|}{0}      \\ \cline{2-4} 
\multicolumn{1}{c|}{}                        & \multicolumn{1}{c|}{Damaged} & \multicolumn{1}{c|}{15}      & \multicolumn{1}{c|}{81}      \\ \cline{2-4} 
\end{tabular}
\caption{Confusion matrix corresponding to sensor node 1}
\label{cf1}
\end{figure}

\begin{figure}[ht!]
\centering
\renewcommand{\arraystretch}{1.2}
\begin{tabular}{llcc}
                                             &                              & \multicolumn{2}{c}{Detected}                                \\ \cline{3-4} 
                                             & \multicolumn{1}{l|}{}        & \multicolumn{1}{c|}{Healthy} & \multicolumn{1}{c|}{Damaged} \\ \cline{2-4} 
\multicolumn{1}{c|}{\multirow{2}{*}{Actual}} & \multicolumn{1}{c|}{Healthy} & \multicolumn{1}{c|}{4}      & \multicolumn{1}{c|}{0}      \\ \cline{2-4} 
\multicolumn{1}{c|}{}                        & \multicolumn{1}{c|}{Damaged} & \multicolumn{1}{c|}{4}      & \multicolumn{1}{c|}{92}      \\ \cline{2-4} 
\end{tabular}
\caption{Confusion matrix corresponding to sensor node 2}
\label{cf2}
\end{figure}

\begin{table}[h!]
\centering
\caption{Performance evaluation of INDDE in lab-scale testbed}
\label{perf}
\renewcommand{\arraystretch}{1.2}
\begin{tabular}{|p{1cm}|p{1.2cm}|p{1.2cm}|p{1.2cm}|p{1.2cm}|p{1.2cm}|p{1.2cm}|}
\hline
Nodes & Accuracy (in \%) & Precision & Recall & F-score & Type-I error & Type-II error \\ \hline
1     &85\%    &0.21    &1   &0.34   &0.15   &0      \\ \hline
2     &96\%    &0.5     &1   &0.66   &0.04     &0    \\ \hline
\end{tabular}
\end{table}

\section{Conclusions}
\label{conc}
This paper presented a robust on-node damage detection mechanism (INDDE) for WSN-based SHM systems by adopting the philosophy of edge computing. INDDE is
well-suited mechanism for real-life applications where data from the damaged state are not readily available to train the model for damage detection.  INDDE was evaluated through two comprehensive case studies considering different damage scenarios. In the first case study on real data-set obtained from Yonghe bridge health monitoring system, INDDE resulted 96-100\% accuracy of correctly identifying the health of the bridge. In the second case-study on lab-scale test set-up, INDDE resulted 85-96\% damage detection accuracy. The experimental results of these case studies demonstrated the ability of INDDE to achieve low false alarm rate. Moreover, INDDE involves no message exchanges which directly benefits in network traffic reduction as well as low energy consumption of the sensor nodes.

\section*{Acknowledgements}
The authors would like to thank Department of Science and Technology (DST), Government of India; Ministry of Science, Technology and Research, Government of Sri Lanka; and University of Peradeniya, Sri Lanka for their support under the Project No. DST/INT/SL/P-15/2016.

\bibliographystyle{unsrt}  
\bibliography{SHM}

\begin{thebibliography}{10}

\bibitem{shm_survey}
Adam~B Noel, Abderrazak Abdaoui, Tarek Elfouly, Mohamed~Hossam Ahmed, Ahmed
  Badawy, and Mohamed~S Shehata.
\newblock Structural health monitoring using wireless sensor networks: A
  comprehensive survey.
\newblock {\em IEEE Communications Surveys \& Tutorials}, 19(3):1403--1423,
  2017.

\bibitem{kaloop4}
Stefania Arangio and Franco Bontempi.
\newblock Structural health monitoring of a cable-stayed bridge with bayesian
  neural networks.
\newblock {\em Structure and Infrastructure Engineering}, 11(4):575--587, 2015.

\bibitem{vibration}
Joan~R. Casas and John~James Moughty.
\newblock Bridge damage detection based on vibration data: Past and new
  developments.
\newblock {\em Frontiers in Built Environment}, 3:4, 2017.

\bibitem{intro}
Alkindi~Rizky Dzulqarnain.
\newblock Distributed processing for operational modal analysis of bridge
  infrastructures using wireless sensor networks.
\newblock Master's thesis, University of Twente, 2017.

\bibitem{shm-cao}
Md~Zakirul~Alam Bhuiyan, Jie Wu, Guojun Wang, and Jiannong Cao.
\newblock Sensing and decision making in cyber-physical systems: The case of
  structural event monitoring.
\newblock {\em IEEE Transactions on Industrial Informatics}, 12(6):2103--2114,
  2016.

\bibitem{shm_framework}
C~Arcadius Tokognon, Bin Gao, Gui~Yun Tian, and Yan Yan.
\newblock Structural health monitoring framework based on internet of things: A
  survey.
\newblock {\em IEEE Internet of Things Journal}, 4(3):619--635, 2017.

\bibitem{shm_fog}
Divya Sinha, Kavish Doshi, and M~Rajasekhara Babu.
\newblock An efficient approach to civil structures health monitoring using fog
  computing as clusters through 5g network environment.
\newblock {\em Advances in Systems Science and Applications}, 18(3):123--143,
  2018.

\bibitem{gda}
Rahul~Kumar Verma, Sourabh Bharti, and Kiran~Kumar Pattanaik.
\newblock Gda: Gravitational data aggregation mechanism for periodic wireless
  sensor networks.
\newblock In {\em 2018 IEEE SENSORS}, pages 1--4. IEEE, 2018.

\bibitem{god}
Sourabh Bharti and Kiran~K Pattanaik.
\newblock Gravitational outlier detection for wireless sensor networks.
\newblock {\em International Journal of Communication Systems},
  29(13):2015--2027, 2016.

\bibitem{edgemining}
Elena~I Gaura, James Brusey, Michael Allen, Ross Wilkins, Dan Goldsmith, and
  Ramona Rednic.
\newblock Edge mining the internet of things.
\newblock {\em IEEE Sensors Journal}, 13(10):3816--3825, 2013.

\bibitem{shm-stat}
Szymon Gres, Martin~Dalgaard Ulriksen, Michael D{\"o}hler, Rasmus~Johan
  Johansen, Palle Andersen, Lars Damkilde, and S{\o}ren~Andreas Nielsen.
\newblock Statistical methods for damage detection applied to civil structures.
\newblock {\em Procedia engineering}, 199:1919--1924, 2017.

\bibitem{decentralized}
Igor~L Santos, Luci Pirmez, Luiz~R Carmo, Paulo~F Pires, Fl{\'a}via~C Delicato,
  Samee~U Khan, and Albert~Y Zomaya.
\newblock A decentralized damage detection system for wireless sensor and
  actuator networks.
\newblock {\em IEEE Transactions on Computers}, 65(5):1363--1376, 2015.

\bibitem{kaloop}
Mosbeh~R Kaloop and Jong~Wan Hu.
\newblock Stayed-cable bridge damage detection and localization based on
  accelerometer health monitoring measurements.
\newblock {\em Shock and Vibration}, 2015, 2015.

\bibitem{ML}
Rafaelle~Piazzaroli Finotti, Alexandre~Abrah{\~a}o Cury, and Fl{\'a}vio
  de~Souza Barbosa.
\newblock An shm approach using machine learning and statistical indicators
  extracted from raw dynamic measurements.
\newblock {\em Latin American Journal of Solids and Structures}, 16(2), 2019.

\bibitem{distributed}
Qing Ling, Zhi Tian, and Yue Li.
\newblock Distributed decision-making in wireless sensor networks for online
  structural health monitoring.
\newblock {\em Journal of Communications and Networks}, 11(4):350--358, 2009.

\bibitem{santosML}
Adam Santos, Eloi Figueiredo, MFM Silva, CS~Sales, and JCWA Costa.
\newblock Machine learning algorithms for damage detection: Kernel-based
  approaches.
\newblock {\em Journal of Sound and Vibration}, 363:584--599, 2016.

\bibitem{liu2012distributed}
Xuefeng Liu, Jiannong Cao, Wen-Zhan Song, and Shaojie Tang.
\newblock Distributed sensing for high quality structural health monitoring
  using wireless sensor networks.
\newblock In {\em 2012 IEEE 33rd Real-Time Systems Symposium}, pages 75--84.
  IEEE, 2012.

\bibitem{forstner}
Ernst Forstner and Helmut Wenzel.
\newblock The application of data mining in bridge monitoring projects:
  exploiting time series data of structural health monitoring.
\newblock In {\em 2011 22nd International Workshop on Database and Expert
  Systems Applications}, pages 297--301. IEEE, 2011.

\bibitem{localized}
Qing Ling, Zhi Tian, Yuejun Yin, and Yue Li.
\newblock Localized structural health monitoring using energy-efficient
  wireless sensor networks.
\newblock {\em IEEE Sensors Journal}, 9(11):1596--1604, 2009.

\bibitem{o2017long}
Sean~M OConnor, Yilan Zhang, Jerome~P Lynch, Mohammed~M Ettouney, and Peter~O
  Jansson.
\newblock Long-term performance assessment of the telegraph road bridge using a
  permanent wireless monitoring system and automated statistical process
  control analytics.
\newblock {\em Structure and infrastructure engineering}, 13(5):604--624, 2017.

\bibitem{2017real}
Osama Abdeljaber, Onur Avci, Serkan Kiranyaz, Moncef Gabbouj, and Daniel~J
  Inman.
\newblock Real-time vibration-based structural damage detection using
  one-dimensional convolutional neural networks.
\newblock {\em Journal of Sound and Vibration}, 388:154--170, 2017.

\bibitem{onesvm}
Ali Anaissi, Nguyen Lu~Dang Khoa, Samir Mustapha, Mehrisadat~Makki Alamdari,
  Ali Braytee, Yang Wang, and Fang Chen.
\newblock Adaptive one-class support vector machine for damage detection in
  structural health monitoring.
\newblock In {\em Pacific-Asia Conference on Knowledge Discovery and Data
  Mining}, pages 42--57. Springer, 2017.

\bibitem{summary}
Jerome~P Lynch and Kenneth~J Loh.
\newblock A summary review of wireless sensors and sensor networks for
  structural health monitoring.
\newblock {\em Shock and Vibration Digest}, 38(2):91--130, 2006.

\bibitem{SHMsystem}
Cristian-Claudiu Comisu, Nicolae Taranu, Gheorghita Boaca, and Maria-Cristina
  Scutaru.
\newblock Structural health monitoring system of bridges.
\newblock {\em Procedia engineering}, 199:2054--2059, 2017.

\bibitem{modal}
Kai-Chun Chang and Chul-Woo Kim.
\newblock Modal-parameter identification and vibration-based damage detection
  of a damaged steel truss bridge.
\newblock {\em Engineering Structures}, 122:156--173, 2016.

\bibitem{rageh}
Ahmed Rageh, Daniel~G Linzell, and Saeed~Eftekhar Azam.
\newblock Automated, strain-based, output-only bridge damage detection.
\newblock {\em Journal of Civil Structural Health Monitoring}, 8(5):833--846,
  2018.

\bibitem{hu-comparison}
Wei-Hua Hu, {\'A}lvaro Cunha, Elsa Caetano, Rolf~G Rohrmann, Samir Said, and
  Jun Teng.
\newblock Comparison of different statistical approaches for removing
  environmental/operational effects for massive data continuously collected
  from footbridges.
\newblock {\em Structural Control and Health Monitoring}, 24(8):e1955, 2017.

\bibitem{li2010statistical}
Weiming Li, Hongping Zhu, Hanbin Luo, and Yong Xia.
\newblock Statistical damage detection method for frame structures using a
  confidence interval.
\newblock {\em Earthquake engineering and engineering vibration},
  9(1):133--140, 2010.

\bibitem{hackmann}
Gregory Hackmann, Weijun Guo, Guirong Yan, Zhuoxiong Sun, Chenyang Lu, and
  Shirley Dyke.
\newblock Cyber-physical codesign of distributed structural health monitoring
  with wireless sensor networks.
\newblock {\em IEEE Transactions on Parallel and Distributed Systems},
  25(1):63--72, 2013.

\bibitem{dist-ssi}
Soojin Cho, Jong-Woong Park, and Sung-Han Sim.
\newblock Decentralized system identification using stochastic subspace
  identification for wireless sensor networks.
\newblock {\em Sensors}, 15(4):8131--8145, 2015.

\bibitem{cluster-shm}
Kun Fang, Chengyin Liu, and Jun Teng.
\newblock Cluster-based optimal wireless sensor deployment for structural
  health monitoring.
\newblock {\em Structural Health Monitoring}, 17(2):266--278, 2018.

\bibitem{wang2007}
Miaomiao Wang, Jiannong Cao, Bo~Chen, Youlin Xu, and Jing Li.
\newblock Distributed processing in wireless sensor networks for structural
  health monitoring.
\newblock In {\em International Conference on Ubiquitous Intelligence and
  Computing}, pages 103--112. Springer, 2007.

\bibitem{khoa2018structural}
Nguyen Lu~Dang Khoa, Mehrisadat~Makki Alamdari, Thierry Rakotoarivelo, Ali
  Anaissi, and Yang Wang.
\newblock Structural health monitoring using machine learning techniques and
  domain knowledge based features.
\newblock In {\em Human and Machine Learning}, pages 409--435. Springer, 2018.

\bibitem{yao2012autoregressive}
Ruigen Yao and Shamim~N Pakzad.
\newblock Autoregressive statistical pattern recognition algorithms for damage
  detection in civil structures.
\newblock {\em Mechanical Systems and Signal Processing}, 31:355--368, 2012.

\bibitem{roy2015arx}
Koushik Roy, Bishakh Bhattacharya, and Samit Ray-Chaudhuri.
\newblock Arx model-based damage sensitive features for structural damage
  localization using output-only measurements.
\newblock {\em Journal of Sound and Vibration}, 349:99--122, 2015.

\bibitem{kiranyaz}
Serkan Kiranyaz, Onur Avci, Osama Abdeljaber, Turker Ince, Moncef Gabbouj, and
  Daniel~J Inman.
\newblock 1d convolutional neural networks and applications: A survey.
\newblock {\em arXiv preprint arXiv:1905.03554}, 2019.

\bibitem{zhou2015}
Qifeng Zhou, Hao Zhou, Qingqing Zhou, Fan Yang, Linkai Luo, and Tao Li.
\newblock Structural damage detection based on posteriori probability support
  vector machine and dempster--shafer evidence theory.
\newblock {\em Applied soft computing}, 36:368--374, 2015.

\bibitem{yu2019novel}
Yang Yu, Chaoyue Wang, Xiaoyu Gu, and Jianchun Li.
\newblock A novel deep learning-based method for damage identification of smart
  building structures.
\newblock {\em Structural Health Monitoring}, 18(1):143--163, 2019.

\bibitem{2019vibration}
Hamid Khodabandehlou, G{\"o}khan Pekcan, and M~Sami Fadali.
\newblock Vibration-based structural condition assessment using convolution
  neural networks.
\newblock {\em Structural Control and Health Monitoring}, 26(2):e2308, 2019.

\bibitem{avci2018wireless}
Onur Avci, Osama Abdeljaber, Serkan Kiranyaz, Mohammed Hussein, and Daniel~J
  Inman.
\newblock Wireless and real-time structural damage detection: A novel
  decentralized method for wireless sensor networks.
\newblock {\em Journal of Sound and Vibration}, 424:158--172, 2018.

\bibitem{farrar}
Charles~R Farrar and Keith Worden.
\newblock {\em Structural health monitoring: a machine learning perspective}.
\newblock John Wiley \& Sons, 2012.

\bibitem{kaloop2}
Shunlong Li, Hui Li, Yang Liu, Chengming Lan, Wensong Zhou, and Jinping Ou.
\newblock Smc structural health monitoring benchmark problem using monitored
  data from an actual cable-stayed bridge.
\newblock {\em Structural Control and Health Monitoring}, 21(2):156--172, 2014.

\bibitem{incontextiot}
Rahul~Kumar Verma, KK~Pattanaik, Sourabh Bharti, and Divya Saxena.
\newblock In-network context inference in iot sensory environment for efficient
  network resource utilization.
\newblock {\em Journal of Network and Computer Applications}, 2019.

\end{thebibliography}

\end{document}